\documentclass{article}
\usepackage{spconf,amsmath,graphicx}
\usepackage{amsfonts}
\usepackage{url}
\usepackage[subtle]{savetrees}

\title{Diversity and Sparsity: A New Perspective on Index Tracking}
\twoauthors
{Yu Zheng}
{Southwestern University of Finance and Economics\\ArrayStream Technologies}
{Timothy M. Hospedales, Yongxin Yang$^*$\thanks{* Corresponding author: yongxin.yang@ed.ac.uk}}
{University of Edinburgh\\}
\begin{document}
%
\maketitle
\begin{abstract}
We address the problem of partial index tracking, replicating a benchmark index using a small number of assets. Accurate tracking with a sparse portfolio is extensively studied as a classic finance problem. However in practice, a tracking portfolio must also be \emph{diverse} in order to minimise risk -- a requirement which has only been dealt with by ad-hoc methods before. We introduce the first index tracking method that explicitly optimises both diversity and sparsity in a single joint framework. Diversity is realised by a regulariser based on pairwise similarity of assets, and we demonstrate that learning similarity from data can outperform some existing heuristics. Finally, we show that the way we model diversity leads to an easy solution for sparsity, allowing both constraints to be optimised easily and efficiently. we run out-of-sample backtesting for a long interval of $15$ years (2003 -- 2018), and the results demonstrate the superiority of the proposed algorithm.
\end{abstract}

\section{Introduction}

The purpose of index tracking is to create an investment portfolio to replicate the performance of a certain market index, e.g., S\&P500. In general, there are two ways to build such a tracking portfolio: full replication and partial replication. 

Full replication is simply to hold all the assets in the same proportions as the market index. It is the most intuitive index tracking approach and provides perfect tracking performance in a frictionless market. However, in practice, it leads to high transaction cost due to large numbers of index constituents, frequently rebalancing, churn in index members, and illiquid assets \cite{Strub2018,BenidisFeng2018}. 

In contrast, partial replication selects a small subset of assets from the index and rebalances at lower frequency (full replication usually require daily rebalancing). This significantly reduces transaction cost, but affects index tracking accuracy. Thus the optimisation problem of partial replication is to compose a small portfolio of assets with minimum index tracking error. This can be seen as involving two sub-problems: asset selection, selecting which subset of assets to hold; and asset allocation, distributing capital among the selected assets. However, for an optimal solution both of these should be tackled jointly \cite{CanakgozBeasley2009,TakedaAkiko2013,Fastrich2014}.

Finding sparse portfolios that replicate an index is a well studied problem due to its importance and broad relevance. The majority of studies look for a sparse portfolio by adding a cardinality constraint on the portfolio, such as $\ell_0$ norm or its variants. \cite{Gotoh2011} provided a nice review on the role of norm constraints. However, a severe problem for theses approaches is that cardinality-based constraints or their variants tend to result in risk concentration. That is, tracking the index by selecting a few assets tends to result in over-exposure to a single industry sector (e.g., banking), thus making the portfolio riskier due to vulnerability to a downturn in that sector. It is well known that a stock portfolio's risk has diversifiable and non-diversifiable components \cite{EvansJohn1968}. Adding a stock to a portfolio generally reduces diversifiable risk only if the portfolio does not yet account for all diversifiable risks. Thus risk minimisation and sparsity are not completely at odds -- constructing a sparse portfolio can be economically rational as not all assets in the benchmark further reduce diversifiable risk. Nevertheless, existing methods for partial index tracking generate portfolios with too much risk as they do not explicitly model portfolio diversity. 

In this paper we therefore study whether we can form a sparse portfolio that accurately tracks the index while simultaneously being \emph{diverse}, thus gaining the benefits of diversity \cite{StatmanMeir1987}. An imperfect answer is to add an $\ell_2$ norm constraint. This can mitigate multicollinearity and thus serve to increase diversity \cite{TakedaAkiko2013}, but does not induce sufficient sparsity to reduce the asset number significantly and does not account for asset inter-dependence. Another solution is to impose the constraint that selects assets (stocks in particular) from different industry sectors. However, this ad-hoc heuristic does not necessarily produce true diversification. For example Apple (consumer electronics) and Corning (optics) are in different sectors but they are highly correlated, as Corning supplies Apple. Thus we aim to design an algorithm that learns the similarity structure from data to achieve diversity. We introduce a learnable similarity matrix $A$ that helps to enforce diversity during optimisation. Most interestingly, we show that the way we introduce diversity uniquely entails an easy way to achieve sparsity through a reweighed  $\ell_1$ norm. 

\section{Methodology}

Practical partial index tracking has three key requirements: (i) The selected portfolio should have minimum error with respect to the true index. (ii) It should be sparse -- composed of a small subset of the full index. (iii) The selected portfolio should minimise risk through diversity. Prior work only addressed the first two of these requirements, while the methodology proposed here will address all three. We start by introducing the index tracking problem in its simplest form, where only tracking accuracy is optimised. We then present our key contribution -- a mechanism to obtain a diverse portfolio. Finally we show how our diversity mechanism also entails an easy solution to the sparsity problem. 

\subsection{Problem Setting}

Index tracking, in its simplest form, is a linear regression problem,
\begin{equation}
\label{base_obj}
\min_w ~ \|Xw-Y\|_2^2
\end{equation}
where $X\in \mathbb{R}^{D\times N}$ are the log-return of assets and $Y\in \mathbb{R}^{D}$ is the target index. $D$ is the number of timesteps (e.g., $D=750$ trading days in three consecutive years), and $N$ is the number of assets (e.g., $N=500$ stocks). $w\in \mathbb{R}^{N}$ is the weight of each asset to hold in order to approximate the index $Y$.

In practice, there are two constraints on $w$: (i) \emph{long only}, which means $w_i \ge {0}, \forall i$ (ii) utilise \emph{all} of the capital, which means $\sum_{i=1}^{N} w_i =1$. Therefore, the objective function becomes,
\begin{equation}
\label{constrained_obj}
\min_{w\ge \mathbf{0},~ \sum_i w_i = 1} ~~ \|Xw-Y\|_2^2
\end{equation}
Eq.~\ref{constrained_obj} is known as a non-negative regression problem with sum-to-one constraint, which can easily be solved by quadratic programming (QP).
\subsection{Diversity}

Diversity is a key property for risk minimisation that has been studied extensively for general portfolio construction problems \cite{WOERHEIDE1992index}. However, it is underused in index tracking. One widely used measure for diversity is $\ell_2$ norm, $\sum_{i=1}^{N}w_i^2$, Under the constraints that $w_i$'s are non-negative and sum-to-one, this is called Simpson diversity index \cite{simpson1949Measurement} in ecology, while it is more commonly known as Herfindahl index in economics. While simple, the key drawback of $\ell_2$ norm is that it does not consider asset inter-dependence. To alleviate this problem, we propose to use,
\begin{equation}
\label{wtAw}
w^T A w
\end{equation}
where $A_{ij}$ is a similarity measure between assets $i$ and $j$, where $0$ means most dissimilar and $1$ means most similar. We have $A_{ii} = 1$ since they are exactly the same asset, and we also assume $A_{ij}=A_{ji}$. We will discuss the choice of $A$ in the following section.

To better understand the role of this term, we can extend $w^T A w$ as,
\begin{equation}
\label{wtAw_extend}
w^T A w = \|w\|_2^2 + 2\sum_{i=1}^{N-1}\sum_{j=i+1}^{N} w_i A_{ij} w_j
\end{equation}
The first term is still the Herfindahl index, but the second term complements diversity, as it discourages buying two assets if they are similar to each other.

One may also build a connection between matrix $A$ in Eq.~\ref{wtAw} and the covariance matrix $\Sigma$ in modern portfolio theory \cite{Markowitz1952PORTFOLIO}. In modern portfolio theory, the term $w^T\Sigma w$ represents the risk (variance) of portfolio, and in our work, $w^T A w$ serves the similar purpose of reducing the risk of several highly correlated assets plummeting simultaneously.

From another perspective, $w^T A w$ is called generalized Tikhonov regularisation \cite{tikhonov1977solutions}. Recall that  common Tikhonov regularisation is simply $\ell_2$ regularisation. Based on the Bayesian interpretation of Tikhonov regularisation, $A$ can be seen as the inverse covariance matrix of $w$.

\subsubsection{Choice of $A$}

A straightforward choice for $A$ is to use asset meta-data. E.g., define $A_{ij}=1$ if asset $i$ (HSBC) and asset $j$ (Citi) are in the same industry sector (Financial services industry), and $A_{ij}=0$ otherwise. In this way, $A$ can be further decomposed as,
\begin{equation}
\label{ZtZ}
A = Z^T Z
\end{equation}
where $Z\in \{0,1\}^{K\times N}$ and $\mathbf{1}^T Z = \mathbf{1}$. $K$ is the number of unique industry sectors, and the $j$th column of $Z$, denoted as $Z_{\cdot,j}$, is the one-hot encoding of the $j$th asset's sector.

Going beyond such heuristics, we ask \emph{can we learn $Z$ from data?} This turns into a clustering problem where $Z_{\cdot,j}$ is the one-hot encoding of the $j$th asset's cluster ID. Arbitrary clustering methods are unsuitable, however, because $X$ is log-return time series data, which tend to be `white noise'. Common  clustering choices, e.g., $k$-means \cite{Lloyd1992least}, are therefore unlikely to work. To this end, we use spectral clustering \cite{Ng2001Spectral} because it provides us the flexibility to define an appropriate similarity measure for this data. 

Note that, it is possible to construct matrix $A$ \emph{without} the decomposable assumption in Eq.~\ref{ZtZ}, but this assumption is helpful in terms of optimisation because it guarantees that $A$ is symmetric positive definite. Furthermore, $Z$ is not necessarily an assignment matrix (asset to cluster). It can be any kind of representations of $X$, but a cluster-assignment representation makes the model easier to interpret. More importantly, building an explicit clustering model is crucial to efficiently realise sparsity as we will see later. However, we do leave the topic of constructing $A$, esp. using a parametrised model like $A=f_\theta(X)$, for future investigations.

\subsubsection{Spectral clustering}

The first step of spectral clustering is to construct an affinity matrix: $S_{ij} = \exp(\frac{-d^2(x_i,x_j)}{\sigma^2})$ if $i \neq j$ and $S_{ii}=0$. $d(x_i,x_j)$ is a distance measure for the $i$th and $j$t column of matrix $X$. The common distance measure is Euclidean distance $d(x_i,x_j)=\|x_i - x_j \|_2$. However since $x_i$'s are log-returns, Spearman's \cite{Spearman1904Proof} or Kendall's \cite{KENDALL1938NEW} rank correlation coefficient is a much better choice because of the robustness. Thus, the distance measure is defined as $d(x_i,x_j)=\sqrt{2(1-\rho(x_i,x_j))}$ where $\rho(x_i, x_j)$ is the rank correlation coefficient.

Then we construct the Laplacian matrix $L=\Lambda^{-\frac{1}{2}}S\Lambda^{-\frac{1}{2}}$ where $\Lambda$ to be the diagonal matrix of which $\Lambda_{ii} = \sum_{j}S_{ij}$. Next, we find the $K$ largest eigenvectors of $L$ (corresponding to the $K$ largest eigenvalues) denoted as $v_1, v_2, \dots, v_K$. Finally, we form matrix $H$ by stacking the eigenvectors in rows, i.e., $H=[v_1^T;v_2^T;\dots;v_K^T]$. For post-processing, we renormalise each of $H$'s columns to have unit length, i.e.,  $H_{ij} \leftarrow \frac{H_{ij}}{(\sum_{i} H_{ij}^2)^{\frac{1}{2}}}$. Finally, we run $k$-means on $H$ (note that each column is an instance).

\subsection{Sparsity}

Sparsity is the crucial propriety of partial index tracking that lowers transaction costs compared to the full index. Thus far we have defined a diversity promoting regulariser, but we have not yet introduced a sparsity constraint. While Eq.~\ref{wtAw} pushes elements of $w$ towards zero, it does not make them sparse. The most common sparsity regulariser is  $\ell_1$ norm, however, it is meaningless in combination with the non-negativity and sum-to-one constraints intrinsic to index tracking. These two constraints mean that $\ell_1$ norm is always $1$ because $|w|_1 = \sum_{i=1}^{N} |w_i|=\sum_{i=1}^{N} w_i=1$.

Our cluster structure introduced earlier provides an elegant solution to this issue. Based on the cluster structure, we can construct a reweighted $\ell_1$ norm \cite{Candes2008Enhancing},
\begin{equation}
\label{l1norm}
\ell_1(w) = \sum_{i=k}^{K} \frac{1}{|\mathcal{C}_i|} \sum_{j \in \mathcal{C}_i} |w_j|
\end{equation}
where $\mathcal{C}_i$ is the set of asset indices in the $i$th cluster, and $|\mathcal{C}_i|$ denotes its size. Eq.~\ref{l1norm} will yield sparsity within each cluster at approximately the same ratio. The vectorized form of Eq.~\ref{l1norm} is,
\begin{equation}
\label{l1norm_vec}
\ell_1(w) = \mathbf{1}^T (Z Z^T)^{-1} Z w
\end{equation}
With Eq.~\ref{constrained_obj}, Eq.~\ref{wtAw}, Eq.~\ref{ZtZ} and Eq.~\ref{l1norm_vec} together, our full objective function can be written as,
\begin{equation}
\label{full_obj}
\begin{split}
\min_w ~~ &\|Xw - Y\|_2^2 + \lambda_1 \| Z w \|_2^2 + \lambda_2 \mathbf{1}^T (Z Z^T)^{-1} Z w \\
&\text{Subject to:}~~ w\ge \mathbf{0}~~ \text{and}~~ \sum_i w_i = 1
\end{split}
\end{equation}
\subsection{Optimisation}

Eq.~\ref{full_obj} can be written as a quadratic programming (QP) problem with both equality and inequality constraint, for which we employ a primal-dual interior-point method \cite{Andersen2003implementing} to solve. The quadratic form of Eq.~\ref{full_obj} is,
\begin{equation}
\label{qp_w}
\begin{split}
\min_w ~~ & \frac{1}{2} w^T P w + q^T w\\
\operatorname{Subject~ to:~} & Gw \le h ~\text{and}~ Aw = b
\end{split}
\end{equation}
where $P=2(X^T X + \lambda_1 Z^T Z)$, $q = \lambda_2 \mathbf{1}^T (Z Z^T)^{-1} Z - 2X^T Y$, $G=-I$, $h=\mathbf{0}$, $A=\mathbf{1}^T$, and $b=1$. Thanks to the design of $A=Z^T Z$ (Eq.~\ref{ZtZ}), we can easily verify that $P$ is symmetric positive definite, which indicates it is also a convex optimisation problem that can be handled by most off-the-shelf QP solvers.

\subsection{Further analysis}

We discuss the role of the second and third term in Eq.~\ref{full_obj}. First, we narrow down to: $\| Zw \|_2^2$. We can rewrite it as $p^T p ~\text{s.t.}~ \sum p_i=1$ where $p_i=Z_{i,\cdot}w$. The physical meaning of $p_i$ is the money that we allocate in the $i$th cluster. By Lagrange multiplier, we can easily tell that $\| Zw \|_2^2$ is minimised when $p_i = \frac{1}{K}, \forall i$. This is very intuitive, because this corresponds to the strategy that we equally allocate the money into every cluster. Second, we analyse the reweighted $\ell_1$ norm term. Similarly, we can rewrite it as $\sum_{i} \frac{p_i}{|\mathcal{C}_i|} ~\text{s.t.}~ \sum p_i=1$, where $p_i$ is again the money that we allocate in the $i$th cluster and $|\mathcal{C}_i|$ is the size of the $i$th cluster. This suggests that, to minimise this term, we need to allocate \emph{all} money for the largest cluster (recall that $|\mathcal{C}_i|$ is a fixed value because $Z$ is given by spectral clustering beforehand). Thus, the second and third term will not agree unless all clusters have exactly the same number of members, which is unlikely in the real world. Therefore, the ratio of $\lambda_1$ and $\lambda_2$ reflects the trade-off between diversity and sparsity.

\section{Experiments}

\subsection{Implementation Details}

Our method has four hyper-parameters: (i) for spectral clustering, there are two: $\sigma$ and $K$; (ii) for the objective function in Eq.~\ref{full_obj}, there are: $\lambda_1$ and $\lambda_2$. Given the scale of experiments, we want to avoid the use of grid search if possible. 

Thus, we set hyper-parameters for spectral clustering by standard heuristic methods. Specifically, $\sigma$ is set by ``median heuristic'' \cite{Gretton2007Kernel}: we first calculate all pairwise distances (excluding self-to-self) and take their median, i.e., $\sigma=\operatorname{median}([d(x_i,x_j),~ \forall~ i\neq j])$. $K$ is set by `` eigengap heuristic'' \cite{Luxburg2007Tutorial}: $K$ is given by the value of $K$ which maximises the ``eigengap" (difference between consecutive eigenvalues), i.e., if we sort all eigenvalues of the Laplacian matrix in an ascending order and the first $K$ eigenvalues are very small, but the $K+1$ one is relatively large.

$\lambda_1$ and $\lambda_2$ are set by grid search: (i) $\lambda_1 \in [1,10]$ and we sample $20$ evenly spaced numbers; (ii) $\lambda_2 \in [800,1000]$ and we sample $200$ evenly spaced numbers. Note that we can not do cross validation here: as the data are real time series, cross validation may result in invalid situations current values are predicted using both previous and future data. Thus, the training-validation split has to strictly follow time.

The last choice is $\rho(\cdot, \cdot)$ which measures the correlation of $x_i$ and $x_j$. As we have discussed, compared to linear correlation, e.g., Pearson's $r$, rank-based correct is a better choice due to robustness. Here we choose to use Spearman's $\rho$ \cite{Spearman1904Proof}.

\subsection{S\&P500 Index tracking}

To evaluate our proposed method in the real world, we track the S\&P500 index using its exact members.
\begin{figure*}[!ht]
\centering
\includegraphics[width=0.245\linewidth]{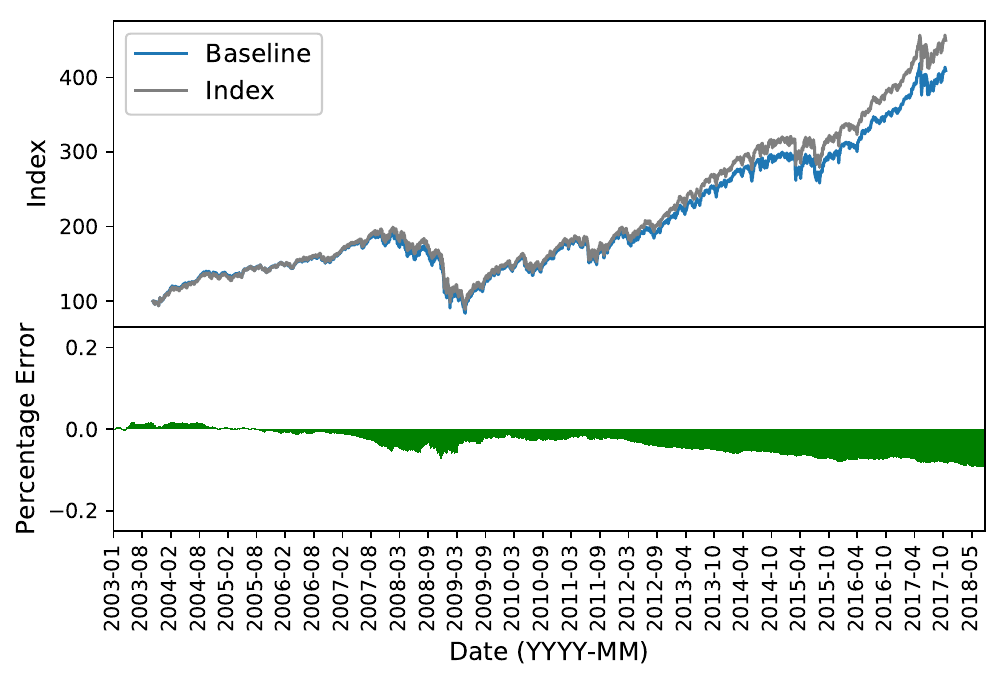}~
\includegraphics[width=0.245\linewidth]{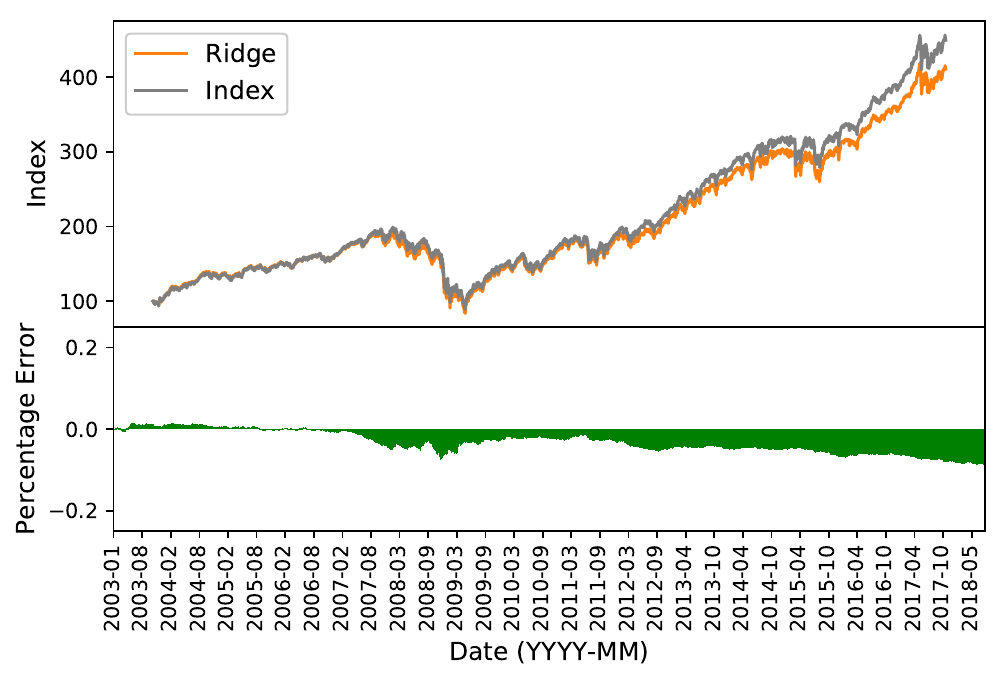}~
\includegraphics[width=0.245\linewidth]{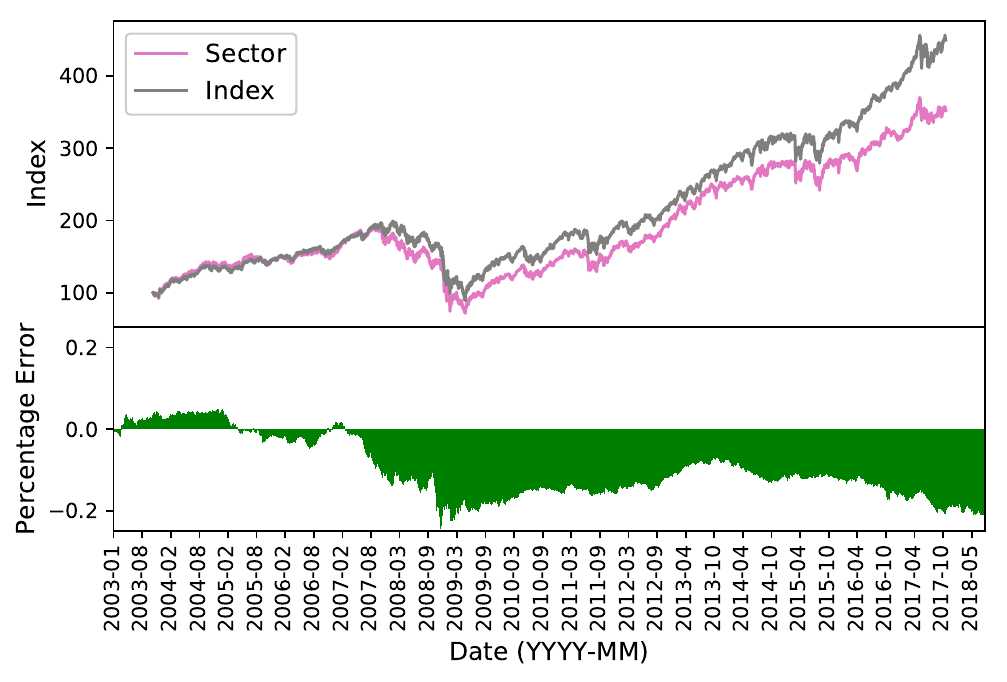}~
\includegraphics[width=0.245\linewidth]{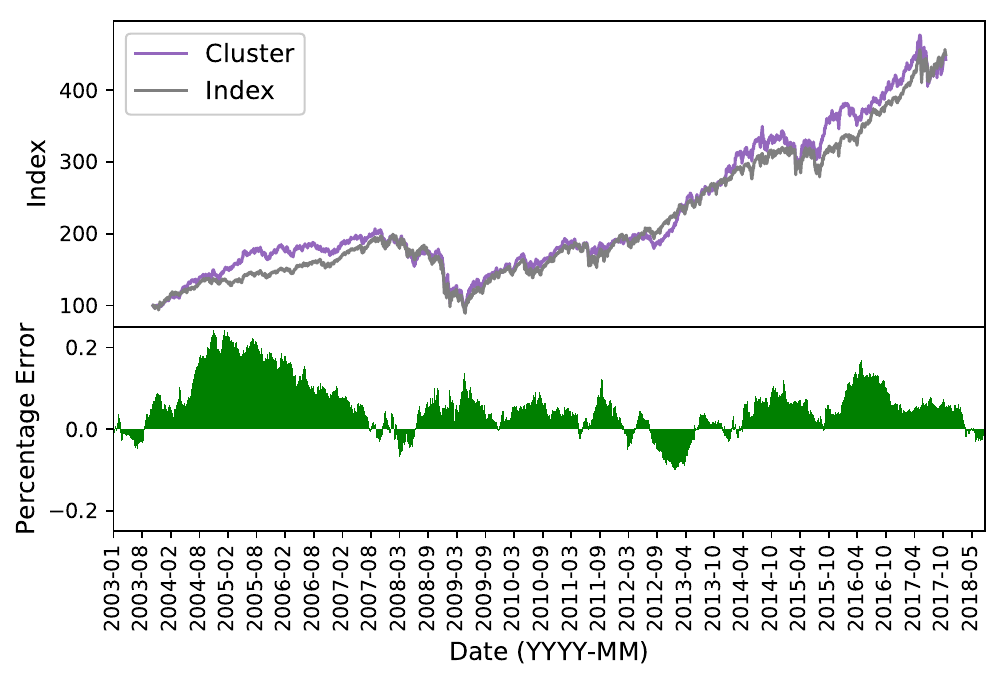}
\caption{Index tracking performance: Top plots are the index and trackers. Bottom is the percentage tracking error $\frac{\hat{y}-y}{y}$.}
\label{fig:num_baseline}
\end{figure*}
\subsubsection{Dataset and settings}

The dataset consists of daily closing prices adjusted for dividends and splits for $852$ stocks from 31 January 2000 to 30 July 2018, a total of $18$ years, provided by The Center for Research in Security Prices (CRSP), which has the most accurate data for security analysis. To avoid the survivorship bias, at each rebalance day, we form the exact constituents of S\&P500 index instead of considering all the $852$ stocks. Furthermore, we also take into account the transaction cost to ensure that our backtesting matches industry practice. We choose the flat-fee pricing model, $\$5.00$ per trade, used by TradeStation, a popular US online stock brokerage firm, to incorporate transaction cost in the backtesting. As the transaction cost is applied on each trade separately, the sparse portfolio will incur less cost compared with the portfolio of a large number of stocks. To enforce the sparsity, we only consider the stocks with weights larger than $10^{-6}$ \cite{ZhangChao2018}. As the transaction cost is related to budget, we assume the initial capital is $\$1$ million in our experiments. Although frequent rebalancing of the portfolio will reduce tracking error, it also entails high transaction cost. To achieve a good balance, we adopt monthly portfolio rebalancing.

\subsubsection{Candidate methods}

We evaluate four methods for the experiment above. \textbf{Baseline}: The objective in Eq.~\ref{constrained_obj}. This is a non-negative regression problem with sum-to-one constraint. This model was proposed in \cite{Meade1989index}.
\textbf{Ridge}: In addition to Eq.~\ref{constrained_obj}, we add an $\ell_2$ norm of $w$. This is known as ridge regression \cite{Hoerl1970Ridge} and its application to index tracking was studied by \cite{DeMiguel2007}. This can also be seen as a reduced version of the proposed method in Eq.~\ref{full_obj} by setting $Z=I$ and $\lambda_2=0$.
\textbf{Sector}: The proposed method in Eq.~\ref{full_obj} where $Z$ is constructed by industry sectors. $Z_{\cdot,j}$ is the one-hot encoding vector that indicates the industry sector of the $j$th stock.
\textbf{Cluster}: The proposed method in Eq.~\ref{full_obj} where $Z$ is constructed by the output of spectral clustering. $Z_{\cdot,j}$ is the one-hot encoding vector that indicates the cluster ID of the $j$th stock.

{Baseline} is hyper-parameter free. {Ridge} has one hyper-parameter which controls the weight of $\ell_2$ norm. {Sector} has two hyper-parameters: $\lambda_1$ and $\lambda_2$. {Cluster} has four hyper-parameters: $\sigma$, $K$, $\lambda_1$ and $\lambda_2$ but we have set $\sigma$ and $K$ heuristically. For those methods that have hyper-parameters, we run extensive grid search to find the best hyper-parameter(s) on the training data.

\subsubsection{Tracking performance}
To evaluate tacking performance, we plot the out-of-sample predictions in Fig.~\ref{fig:num_baseline}. There are two issues to study in tracking performance. First is tracking \emph{accuracy}, as all methods are aspiring to track the index with low error. Baseline, Ridge, and Cluster have similar accuracy, while Sector is slightly worse. Second is the sign of the error: trackers aim to match or exceed the index, and avoid underperforming it. This is affected by sparsity and diversity, where balancing these two is the key challenge. The Ridge approach is low-risk/high-diversity, but underperforms due to incurring high transaction cost for holding the full index. Sector maintains good sparsity, but is insufficiently diverse. Our Cluster approach, comes closest to matching the index due to effective joint optimisation of diversity and data-driven sparsity. To quantitatively evaluate these methods, we calculate the statistics of absolute percentage errors for different methods in Tab.~\ref{tab:sape}, which is corresponding to the integral of green bars in Fig.~\ref{fig:num_baseline}. While the sum/mean directly reflects the tracking accuracy, for which Ridge has the smallest error, we are also interested in which contribute to the sum: the positive error (area above zero) is more tolerable since it means better returns compared to market. Taking this into account, Cluster has the best overall performance.

\begin{table}
\centering
\begin{tabular}{c c c c c}
\hline 
Method & Negative & Positive & Sum & Mean \\
\hline 
Baseline & 145.35 & 5.36 & 150.71 & 3.86\%\\
Ridge & 131.56 & 5.28 & \textbf{136.84} & \textbf{3.51}\%\\
Sector & 397.22 & 16.69 & 413.91 & 10.61\%\\
Cluster & \textbf{21.42} & \textbf{237.17} & 258.59 & 6.63\%\\
\hline 
\end{tabular} 
\caption{Absolute percentage errors for different methods}
\label{tab:sape}
\vspace{-0.7em}
\end{table}

\section{Conclusion}

We presented an elegant model for the index tracking problem that jointly optimises both diversity and sparsity. It is very easy to solve as a standard QP problem, yet achieves excellent performance for both tracking accuracy and the number of stocks traded. It can be seen as a general solution that brings $\ell_1$ norm back into the game for regression problems with non-negativity and sum-to-one constraints when a sparse solution is desired. In future work, we will investigate if it is possible to integrate the ``offline" clustering step into the optimisation problem by exploring options for constructing $A$ or $Z$ matrix end-to-end.

\noindent\textbf{Acknowledgement:} This work has been supported by the Financial Innovation Center of the Southwestern University of Finance and Economics and the Key Laboratory of Financial Intelligence and Financial Engineering of Sichuan Province.\\

\noindent\textbf{Disclaimer:} All authors are faculty. Neither graduate students nor small animals were hurt while producing this paper.

\end{document}